\documentclass[doublecol,linenumbers]{epl2} 
\usepackage{amsmath,amsfonts,amssymb,amsthm}
\usepackage{lineno,hyperref}
\usepackage{marginfix}

\title{Enhanced resolution in Argon and Neon spectra using a Super-Resolution algorithm}
\shorttitle{Super-Resolution algorithm for the Argon and Neon spectra}

\author{L. M. Hoyos-Campo\inst{1} \and A. M. Juarez\inst{1} \and A. Capella\inst{2}}
\shortauthor{F. Author \etal}

\institute{                    
  \inst{1} Instituto de Ciencias F\'isicas -  Universidad Nacional Aut\'onoma de M\'exico, PO Box 48-3, 62251, Cuernavaca, Mor., M\'exico\\
  \inst{2} Instituto de Matem\'aticas - Universidad Nacional Aut\'onoma de M\'exico, Circuito Exterior, C.U., 04510, M\'exico D.F., M\'exico}
	
\pacs{32.30.-r}{Atomic spectra}
\pacs{07.05.Kf}{Data analysis: algorithms and implementation; data management}
\pacs{52.38.-r}{Laser-plasma interactions}

\abstract{This paper presents the principles and application of a super-resolution (SR) technique, based on a L1-Norm minimization procedure. In particular, the technique is applied to improve low-resolution resolution spectra as obtained from the optogalvanic effect in Neon and Argon discharges over the $413$-$423$ nm wavelength range. By applying the super-resolution algorithm to the experimental data, a surprising $70$-fold reduction of the linewidth is achieved allowing to resolve prior indistinguishable peaks. It is found that further improvements on the resolution are limited by the   signal to noise ratio of the original spectra. The importance of a suitable mathematical representation of the experiment and the discussion on other applications of this super-resolution technique in spectroscopy are also presented.}

\begin{document}

\maketitle

\section{Introduction}

In atomic and molecular spectroscopy it is desirable to resolve relevant 
spectroscopic features present is the spectra by achieving the highest possible 
resolution. Unfortunately, the light sources and instruments employed in experiments induce limits in the  final resolution of the measurements.
To tackle this limitation experimentalist traditionally resort to line-stabilized, high resolution light sources along with Doppler free spectroscopy techniques \cite{Paul2010} or supersonic sample cooling \cite{Metcalf2012}. These techniques present, however, the inconvenience of involving relatively elaborated experimental arrangements. 
Here, we present a novel mathematical technique aimed at helping experimentalists to improve the spectral resolution of their data. This technique, called Super-Resolution, is based on the mathematical tools developed within the scope of Compressed Sensing (CS) \cite{CandesRombergTao06,CandesWakin2008}. 
The basic idea behind CS is that,  under the {\it a priori} knowledge that a signal or an image has a sparse representation (see below), 
this can be reconstructed from far fewer data/measurements as  compared  to those required  by the well-known Shannon/Nyquist sampling theory \cite{Candes06-2}. 
Here, the term sparsity means that there exists a basis in the space of functions (e.g. a Fourier basis or Wavelets) where the signal can be represented by a small number of elements.
In spectroscopy, a spectrum of a single element or molecule can be represented as a linear combination of single peaks of a similar shape centered at their specific transition wavelength values.
In this case, the spectrum can be represented by a small number of parameters, namely the locations of the transitions, their corresponding widths, and their amplitudes. 
These parameters constitute the sparse representation in this case.\\
In a general context, Compressed Sensing can be regarded as an optimal sampling and 
interpolation algorithm for sparse signals in the appropriate base. 

Along the ideas of CS, some new versions of algorithmic or computational Super-Resolution (SR) algorithms 
have been developed recently. These SR algorithms have  been successfully applied to many practical cases, including 
medical imaging \cite{Hayit}, satellite imaging, and video applications
(for instance sec. 1.2 in \cite{FocautRauhut2010} and  the references within). 

In the context of the present work, SR algorithms are applied to the particular case of recovering high resolution features such as peak centers, from  experimental spectra. We show here that this kind of mathematical problem  can be treated by the recently developed mathematical theory of Super-Resolution by Cand\`es and Fernandez-Granda \cite{Candes,Candes2}.

The aim of this paper is to show that SR algorithms can  also be a valuable tool 
in spectroscopy by allowing to resolve  peaks that cannot be separated from 
the original data due to instrumental limits in spectral resolution \cite{Dror,Pavel}.
In particular, we show here that the SR algorithm can dramatically 
improve the resolution of measurements, in particular allowing a 
surprising effective $70$-fold reduction in the linewidth of the original spectral measurements. 
This improvement allows to separate apart real structures that are not resolved in the original experimental data. 
In particular, in order to test and quantitatively show the power of the SR technique, we present here 
experimental optogalvanic spectra of well-known Neon and Argon atomic transitions. Upon the application of the SR technique we can quantify dramatic improvements in the resolved features. 
%
%
The optogalvanic effect (OGE) consist of a change in the total current of a plasma when interacting with laser light. The change is resonant when the laser wavelength corresponds to an atomic or molecular transition of the gas species present in the plasma \cite{Barbieri}.
Although the specific example shown here is devoted to the OGE effect, we believe SR algorithms can be applied in other spectroscopic areas, including photoionization, photoabsorption, Raman or others. 
To the best of our knowledge this is the first application of super-resolution  algorithms based on compressed sensing ideas aimed at improving  the resolution of spectroscopic measurements.

\section{Super-Resolution algorithms}\label{Superresolution}

In any imaging system, the resolution limit is measured by the shortest separation between two features,  before they mingle into one. The limitations of the instrument used and, on a more 
fundamental level, the diffraction limit set the constraints to the maximum experimentally achievable resolution in any optical system. There are currently several instances  and many different approaches to SR in imaging, and we invite the reader to revise the exciting applications of optical super-resolution currently being developed(e.g.  \cite{Park2003} and references within). 
In an analogous manner as in optics, in any type of optical spectroscopy, the spectral resolution is limited by the linewidth of the light source and quantified by the minimum separation in wavelength required to  clearly separate apart two different transitions. This has been traditionally considered the ultimate limit of resolution in spectroscopy.  
Since the applications of SR techniques to spectroscopy are scarce and not widespread in the experimental community, we will give a basic introduction in the context of  optogalvanic spectroscopy.   

Let us denote by $y$ a one dimensional  acquired signal as obtained from a source $x$ through 
an optical or any other measuring device. In our particular case $y$ corresponds to the optogalvanic signal as and $x$ corresponds to the original spectrum as produced by the physical  interaction of the laser with the plasma. Assume, as it happens in our optogalvanic case, that 
the values of $y$ and $x$  both depend on the wavelength $\lambda$. A mathematical model 
of such a process can be described by the following convolution model
\begin{equation}\label{conv}
y(\lambda) = \int_0^\infty G(\lambda-s) x(s) ds + w(\lambda)
\end{equation}
Here, $G$ represents, mathematically,  the effect of the measurement equipment/process upon the original physical data, $s$ is an integration variable and $w$ is a noise term that can be deterministic or stochastic and such that
\begin{equation}\label{noise}
\Vert w\Vert_{L^1}=\int_0^\infty |w(\lambda)|d\lambda <\varepsilon,
\end{equation}
but otherwise arbitrary. Here, $\Vert w\Vert_{L^1}$ denotes the continuum version of the vector norm, colloquially known as the taxicab norm. Notice that $\varepsilon\ge 0$ is a constant measuring the noise level of the acquired signal. In the reciprocal space \eqref{conv} becomes
\begin{equation}\label{reciprocal}
\hat{y}(\nu)=\left\{
\begin{array}{ll}
\widehat{G}(\nu)\widehat{x}(\nu) + \widehat{w}(k) & \text{ for } k\in [-\nu_c,\nu_c]\\
\widehat{w}(\nu) &\text{otherwise }
\end{array}
\right.
\end{equation}
where $\widehat{G}$ denotes the Fourier transform of the function $G$ and $\nu=1/\lambda$ 
is the usual definition of wavelength per unit distance. In this mathematical representation the
instrument $G$ can be seen as a low pass filter that limits the bandwidth of an otherwise  ideal 
and spectrally narrow source signal.  In \eqref{reciprocal} $\nu_c>0$ is the cutoff number of  
wavelength per unit distance of such a filter. For the moment, and for simplicity in the 
exposition, we discard the noise term of \eqref{noise} by letting $\varepsilon=0$. 

Formally, the aim of the SR algorithms is to find an estimate $x_\text{est}$ of the source $x$ such that:
\begin{equation}\label{SRproblem} 
\left\{\begin{array}{l}
\text{(i) }\  \widehat{G* x_\text{est}}=\widehat{G* x} \quad\text{ on } [-\nu_c,\nu_c] \\
\text{(ii) }\  \widehat{x}_\text{est} \ \text{ is supported in } \ [-\nu_\text{hi}, \nu_\text{hi}]
\end{array}\right.
\end{equation}
where $\nu_\text{hi}> \nu_c$ and $ \widehat{x}_\text{est}$ denotes the Fourier transform of the
estimate $x_\text{est}$. In simpler terms, the aim of SR is to use the frequency limited information 
as produced by the instrument (or light source) to make an extrapolation (in this case $x_\text{est}
$) to higher frequencies in reciprocal space. This extrapolation would allow, upon an inverse
Fourier transformation, to obtain a far richer spectrum, i.e. the ``super-resolved" spectrum as 
compared to that obtained originally from $y$, which corresponds to the measured spectrum. 
A quantitative measure of the improvement in our signal can be defined in terms the 
super-resolution factor $SRF=\nu_\text{hi}/\nu_{c}$. An alternative way to measure the 
effect of the SR algorithm on the data is to define  the  percentage improvement of
 the signal (PEIS) as $(1-(SRF)^{-1})*100$. 
It is well known that the problem given as  in \eqref{SRproblem} is hopelessly ill posed, namely 
the solution $x_\text{est}$ may not exist or may not be unique~\cite{Slepian}. This means that, based 
on a truncated Fourier transform, one may find a very large set of extrapolations that faithfully 
represent the original data  within the interval $[-\nu_c,\nu_c]$, but which are wildly different 
outside this interval. This many solutions would not necessarily have any physical meaning 
even if they satisfied \eqref{SRproblem}.
Remarkably, in the case where the source $x$ is composed of a ``well separated" train of 
point--like sources,  problem \eqref{SRproblem} becomes tractable. This kind of sources can be modeled by
\begin{equation}\label{trains}
x=\sum_{j\in T} a_j\delta_{\lambda_j},
\end{equation}
where ${\lambda_j}$ and $a_j$ are the locations (in the spectrum)  and intensities of the point sources, respectively, $\delta_\tau$ is the  Dirac delta function\footnote{In mathematical rigor  $\delta_\tau$ is a Dirac measure.} and $T$ is a given set of numeration indexes.  

As mentioned in the introduction, the above model can be used as a simple model for atomic or 
molecular transitions. In these,  their linewidths even when Doppler broadened,  
are very narrow as compared to the spectral resolution of the laser used in the experiment. We emphasize again that the models and method in this paper will be only useful to determine 
the locations and relative intensities of  the spectral lines. The present analysis is not valid when the 
linewidth of the laser is small as compared to the  Doppler broadening of the transitions. For 
such a case a more realistic modeling of the transition linewidths  on the source should be proposed, but that goes beyond the scope of this work. 

Next, we define the {\it minimum  separation} of the point sources as $\Delta=\min_{i,j\in T}|\lambda_j-\lambda_i|$ where $\lambda_i,\lambda_j$ denote again their locations on the spectral line. The main result of the SR theory  states that, if the point sources are sufficiently apart from each other, namely
\begin{equation}\label{minsep}
\Delta \ge   2/\nu_c, 
\end{equation}
then there  exist an efficient algorithm  that recovers exactly the original source.  In other 
words, problem \eqref{SRproblem}  has  a unique solution $x_\text{est}=x$. This solution  can be 
computed by solving  a convex minimization problem  \cite{Candes2} that is, finding the 
minimum of a convex function under constrains (e.g. \eqref{discrete} below).  
Notice that there is no bound on $\nu_{hi}$ so in principle, at least mathematically, the improvement on the resolution may be infinite (namely, $SRF=+\infty$), and the 
location of the point sources can be determined with infinite precision.  As we will see latter,
 the presence of noise imposes natural constrain on $SRF$, hence on the improvement 
 that can be obtained by the estimate $x_\text{est}$. 
Before stating the SR algorithm in detail, we need another ingredient. In digital signal 
processing, as in any digital data acquisition experiment,  there is a finite size in the 
wavelength step that is acquired, so sources and signals can be regarded both as 
discrete finite vectors $x=(a_1,\dots,a_{N})$ and $y=(y_1,\dots,y_{N})$ measured at 
fixed grid of equally spaced wavelengths $\lambda$.

\begin{figure}[h!]
  \centering
    \includegraphics[scale=0.9]{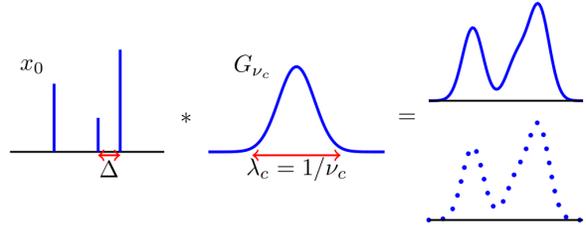}
  \caption{Convolution of a train of point sources with a Gaussian filter. The graphs on the right show:(Top) the results for a continuous signal. (Bottom) the same signal after sampling on a discrete grid. The basic features of the continuous signal can be recovered by an interpolation of the discrete one. Hence, regarding the filter $G$, both graphs have 
the same cutoff wavelength per unit distance $\nu_c$.}
  \label{fig_filter}
\end{figure}

In the case of continuous signals, the filter resolution is defined by $\nu_c$, 
that is, the width of the low-pass filter $G$ (see Figure~\ref{fig_filter}) and it is related to the physical properties of the experimental setting. In the experiments we also have 
another loss of resolution due to the finite size of sampling , i.e. the wavelength step. Notice 
that the latter loss of resolution may change, for example, by reducing the scanning step of the laser used in the experiment,  but 
this does not affect the resolution of the filter $G$. The aim of SR algorithms is to reverse 
(or deconvolve) the effect of the filter $G$ on the source signal. Once the sampling step 
size is smaller that certain threshold its effect on the SR algorithms is negligible.  This 
important difference, combined with the unboundedness of the $\nu_\text{hi}$  
and the separated point source nature of the source signal explain 
why, in our case, the improvement in resolution seems to be independent of the finite 
sampling step size (see Figure~\ref{fig_filter}).
Now, again discarding noise, \eqref{conv} becomes   
\begin{equation*} 
\begin{split}
y_\ell=\sum_{j=1}^{N} G\left(\lambda_\ell-s_j\right) a_j,  \quad  |\lambda_\ell & |\le \nu^{-1}_c \\
& \quad \text{ for } \ell\in\{0,\dots,N-1\},
\end{split}
\end{equation*}
or in more compact notation we write $y=Gx$ where $G$ is an $N\times N$ matrix whose entries 
are given by $G_{\ell j}=G\left(\lambda_\ell-s_j\right)$.
Hence, the SR problem \eqref{SRproblem} is reduced to finding out the entries $a_j$ of the vector $x$. 
As proposed in \cite{Candes2}, to solve this problem it is enough to solve a convex minimization 
program, namely 
\begin{equation}\label{discrete}
\left\{\begin{array}{l}
\displaystyle \min_{\tilde{x}} \Vert \tilde{x} \Vert_{L^1}\\
 \ \displaystyle \text{subject to } G\tilde{x}=y
\end{array}\right.
\end{equation}
where $\Vert \tilde{x} \Vert_{L^1}=\sum_{j=1}^{N} |a_j|$.  The main result in \cite{Candes2} is that 
the problem \eqref{discrete} has a unique  solution $x_\text{est}$  that coincides with the original $x$ 
over the original range of inverse space, but has a much larger span of harmonics beyond the interval defined by the original cutoff wave number. This is a surprising 
result and problem \eqref{discrete} is at the core of the  mathematical theory of super resolution. 
The main feature of the $L^1$-minimization, that ``explains why" the solution of \eqref{discrete} 
recovers the original signal, is that, it is an efficient way to detect the support of the sparsest
signals (that is, signals that contain the least number of peaks)  among all possible signals that 
produce the same observable data $y$ via the filter $G$. The role of the minimum separation 
condition \eqref{minsep} is to make the solution of problem \eqref{discrete} unique and hence it must coincide 
with the original source signal. Other important feature of problem \eqref{discrete} is that it is a 
convex optimization problem under linear constrains. This is a standard problem in non-linear 
programming  that even some spreadsheet software is able to solve. In the presence of noise, that is for $\varepsilon >0$ above, the reconstruction given by 
\eqref{discrete} will not be exact \cite{Candes}, but should instead satisfy 
\begin{equation}
\Vert G(x_\text{est}-x) \Vert_{L^1} \le C_0 (SRF)^2 \varepsilon,
\end{equation}
where $C_0>0$ is a positive constant. Therefore, the quality (or accuracy) of the estimated 
vector $x_\text{set}$ degrades linearly with the noise and quadratically with the SRF. Hence, 
there is a tradeoff between the noise level and the SRF factor that limits the performance of 
the SR algorithm. Is important to emphasize that, in order to apply this algorithm successfully, 
a good control on the signal to noise ratio is required in the original experimental data.  

\section{Filter model in optogalvanic spectroscopy}\label{FilterOG}

The last ingredient that is needed to apply the above SR algorithm \eqref{discrete} to any 
experimental setting is to identify filter $G$ and estimate its parameters.  In general, the 
shape of the optogalvanic peaks does not present a unique profile condition and it may depend, 
for example, on the current of the plasma \cite{Bachor1982}. 
Nevertheless, it is known that the upper part of each peak can be well approximated at first 
order by a Gaussian profile, and the lower part by a Lorentzian. Since the aim of our algorithm 
is to identify the position and relative strength of the peaks and these positions are well correlated 
with the upper part of the peaks,  we propose, in a heuristic way, the following Gaussian form of 
the filter
\begin{equation}
\widehat{G}(\nu)=B\exp(-\frac{\nu^2}{a^2}), \quad\text{for } |\nu|<\nu_c
\end{equation}
where $B$ is a normalization constant and $a$ corresponds to the effective 
broadening of the experimental peaks. Note that it is convenient to write $G$ in terms of its 
Fourier transform since in this case it is a diagonal operator (recalling that the Fourier 
transform of a Gaussian is also a Gaussian). What remains is simply to estimate the 
parameters $\nu_c$, $a$ and $B$ that more closely models the experiment. Notice that $\nu_c$ is 
the cut-off frequency and it is an important limit since it represents the frequency (in Fourier or reciprocal 
 space) where the noise to signal ratio is small enough to consider the signal consisting of pure noise. 

\begin{figure}[h!]
  \centering
    \includegraphics[scale=0.362]{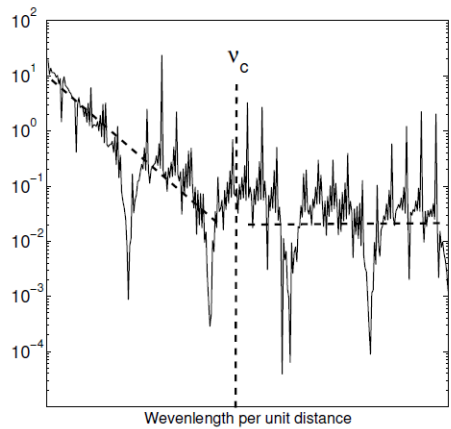}
  \caption{Plot of wavenumber per unit distance $\nu$  vs. $\log|\widehat{y}_\text{fit}/\widehat{x}_\text{know}|$.
  At the cut off wavenumber $\nu_c$ the graph shows a change of behavior.}
  \label{fitting}
\end{figure}

In order to estimate the filter, we propose here a simple but effective approach:
 given the above functional form for the filter, we proceeded to obtain the filter parameters using high resolution 
experimental data $x_\text{know}$ from a gas species. In our present case these correspond to  Neon optical transitions as reported by NIST \cite{NIST} and its corresponding spectrum as measured by our apparatus $y_\text{fit}$.
The procedure to obtain the parameters of the filter is started  by, first, computing the Fourier transform of the measured spectrum and the NIST  
reference spectrum. Second, since in Fourier space the action of filter is multiplicative
 (see \eqref{reciprocal}), the filter can be thus be obtained as the quotient of the  above quantities. That is
\begin{equation}
\widehat{G}(\nu)= \widehat{y}_\text{fit}(\nu) / \widehat{x}_\text{know}(\nu)
\quad\text{ for } \nu\in[-\nu_c,\nu_c],
\end{equation}
where for $\nu\notin [-\nu_c,\nu_c]$ the fraction $\widehat{y}_\text{fit}(\nu) / \widehat{x}_\text{know}(\nu)$ 
only contains the wave number information coming from the noise term (see figure~\ref{fitting}).  To explain 
our fitting method to the experimental data of Neon, in figure~\ref{fitting} we show a plot of the discrete 
cosine transform\footnote{Fourier and cosine transforms are equivalent in this case, here we use the 
latter because it has the advantage of letting us to work with real signals only.} of 
$\log |\widehat{y}_\text{fit}/\widehat{x}_\text{know}|=\log \widehat{G}$ for 
some of our Neon and NIST data on a particular interval. 
First, starting at $\nu=0$ there is 
a noisy signal mounted on a smooth decreasing curve that after certain point $\nu_c$ can be regarded as 
pure noise. The point where the signal behavior changes becomes the estimated cutoff frequency  
 $\nu_c$. 

Once $\nu_c$ is known, the parameters $a$ and $B$ can be determine by a linear regression procedure for 
$\log(|\widehat{y}_\text{fit}/\widehat{x}_\text{know}|)$ on the interval $(0,\nu_c)$. 
To have a more robust estimation, the determination of $\nu_c$ and the linear regression is repeated on intervals around different peaks and then the reported values are given by their average value. This approach to estimate the parameters is effective and good enough for the current application. 


\section{Experiment}

\begin{figure}[h!]
  \centering
    \includegraphics[scale=0.4]{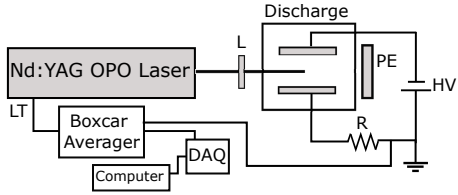}
  \caption{Experimental arrangement. L:PlanoConvex lens f$=150$ mm, PE: Pyroelectric sensor, R: ballast resistor $3.9$ kOhms, HV: DC power supply, LT: laser trigger.}
  \label{Diagrama}
\end{figure}

The experimental arrangement employed is shown in figure \ref{Diagrama}. It consists of three main parts: the light source, the plasma discharge, and the detection part. 
The electrodes, of planar geometry and with a separation gap of $7$ mm, were made of copper. The Argon and Neon used was of research grade purity. The vacuum system consists of a Pfeiffer turbomolecular pump of $100$ l/s that allows us to reach a base pressure of $10^{-5}$ Torr. Pressure was measured by using a MKS baratron capacitive manometer type $626$ and a Varian Eyesys mini-BA vacuum gauge. We use a DC power supply $HP6516$ with a voltage range of ($0-3000$) Volts.\\ 
To excite the plasma we used an Ekspla Optical Parametric Oscillator (OPO) pumped by a Nd:YAG pulsed laser. For the optogalvanic effect  the voltage variations were recorded using a ballast resistor of $3.9$ KOhm. This voltage was amplified by a SR$280$ StanfordResearch Systems boxcar averager. The power of the laser was measured using an Ophir pyroelectric sensor PE$10-$S. These measurements were send to a National Instruments USB$-6009$ Data Acquisition System (DAQ) and collected in a computer using a LabVIEW automatization program.
The laser resolution in the visible range reported here was $0.1$ nm and its wavelength was varied in an automated way. For each wavelength we collected several voltage values  and the values reported here are an average of at least 20 of these.

\section{Results and discussion}

\begin{figure}[h!]
\centerline{\includegraphics[scale=0.6]{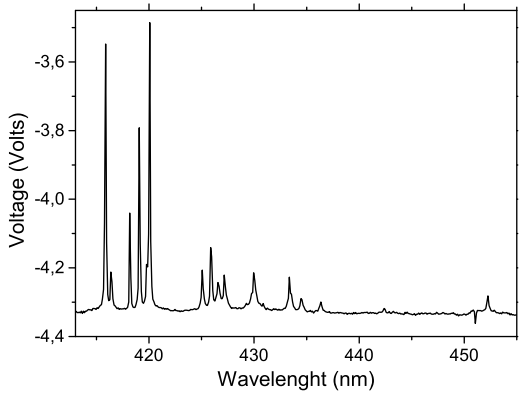}}
\caption{Measured Argon optogalvanic spectra at $3.9$ Torr and a current of $1.21$ mA.}
\label{ArgonOGE}
\end{figure}

\begin{figure}[h!]
   \centerline{\includegraphics[scale=0.6]{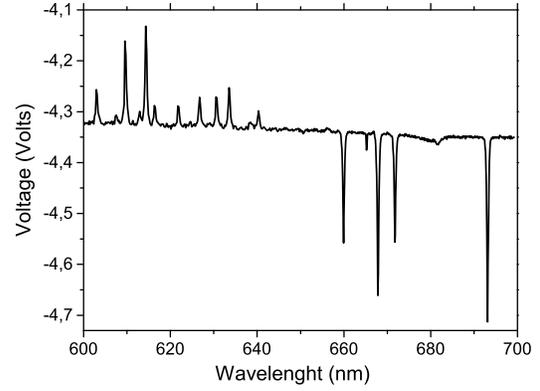}}
   \caption{Measured Neon optogalvanic spectra with a pressure of $3.3$ Torr and a current of $4.08$ mA.}
  \label{NeonOGE}
\end{figure}

The Argon discharge in figure \ref{ArgonOGE} was obtained with a step wavelength variation of $0.1$ nm.
For the Neon discharge in figure \ref{NeonOGE} the wavelength variation was in steps of $0.1$ nm for the ($600-650$)nm wavelength range and $0.2$ nm for the ($650-700$)nm range.\\ 
For the filter parameters estimation we used four different wavelengths ranges (windows) for the  Neon 
experiment and one for the Argon experiment. The latter was used to double check that the main features of the filter 
were the same for both experiments.
The high resolution experimental signal data  used to calculate the filter parameters corresponded 
to the transitions reported in the NIST atomic database for Neon\cite{NIST}. This was used as the 
know source signal $x_\text{know}$. The windows were chosen in such a way that 
some clearly defined peaks of the source signal $x_\text{know}$ were contained inside a single peak in the measured signal $y_\text{fit}$. In table~\ref{table} we show the Neon windows considered, the values of the cutoff wavenumber $\nu_c$ and filter parameter $a$. 
\begin{table}
\centering
    \begin{tabular}{  c | c | c }
    \hline
     Window(nm)& $\nu_c\times10^{8}$ ($cm^{-1}$) & a $\times10^{8}$ ($cm^{-1}$)\\ \hline
     $608.5-610.5$ & $14.0$ & $6.85$ \\ 
     $620.0-623.0$ & $14.7$ & $6.98$ \\ 
     $625.0-628.0$ & $14.3$ & $6.80$ \\ 
     $632.0-634.5$ & $14.5$ & $6.82$ \\
    \hline
    \end{tabular}
    	\caption[Table]{Filter parameter estimation with Neon data.}
	\label{table}
\end{table}
The mean values for this experiment are $a=6.87\times 10^{8} cm^{-1}$ and 
$\nu_c=14.37\times10^{8} cm^{-1}$.  The considered window for Argon was $414.5-417.5$ nm 
and the obtained parameters are $a=6.70\times 10^{8} cm^{-1}$ and $\nu_c=14.7\times 10^{8} cm^{-1}$, 
these values are within 2\% and 5\% correspondence with the Neon mean values, respectively. 
As explained in the previous sections, the resolution of the estimated signal $x_\text{est}$ is 
given by $\nu_\text{hi}=1/\lambda_\text{hi}$, and the SR algorithm does not prescribe any 
bound on $\nu_\text{hi}$ (or on the smallness of $\lambda_\text{hi}$), therefore in principle 
it can be chosen at will. In practice, this is not possible due to the signal to noise ratio and the
numerical stability in the solution of \eqref{discrete}. The latter problem is a consequence of the 
fact that,  for small $\lambda_\text{hi}$ it is necessary to solve a large linear system of $\sim 
1/\lambda_\text{hi}$ equations that may be numerical ill conditioned. 
The numerical scheme of the SR algorithm and the adjustment of the filter parameters was 
performed as follows: First, all signals, $x_\text{know}, y_\text{fit}$ and $y$, were linearly 
interpolated on to a fix grid equally spaced of step size $\lambda_\text{hi}= 10^{-4}$ nm, 
which yields $\nu_\text{hi}= 10^{12}$ cm$^{-1}$. Second, the filter parameters were estimated 
by the above procedure, and third, the minimum in problem \eqref{discrete} was found to obtain $x_\text{est}$. 
There is a final consideration to be made regarding a source of error that originates 
from the experimental offset in the measured intensity of the peaks produced by the boxcar amplifier. 
This error is assimilated in the model of the noise 
term  $w$ in \eqref{conv} as an additive constant. By adjusting this offset, by means of another additive 
constant, the performance of the SR algorithm improves dramatically. In the presented experiments this 
value was set so as to lower the signal by $1.6$ \% with respect to the highest peak (near $420$ nm). 
The numerical implementation was based in the useful Matlab code from \cite{CandesRombergl1magic}. 
\begin{figure}[h!]
   \centerline{\includegraphics[scale=0.6]{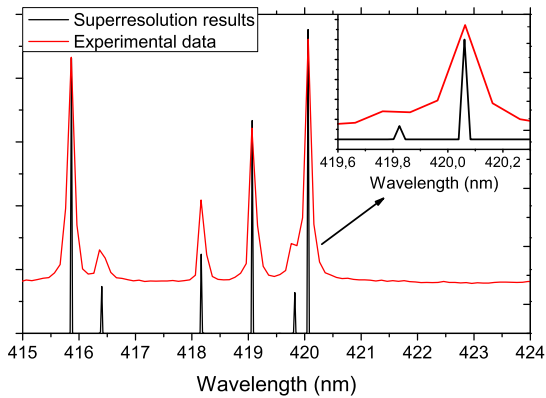}}
  \caption{Experimental Argon optogalvanic spectra and Argon SR results. The spectral resolution improvement allows to resolve some prior experimental indistinguishable peaks.}
  \label{ArgonVsSR}
\end{figure}
In the figure \ref{ArgonVsSR} we show a comparison between the Argon optogalvanic spectra before and after applying the SR algorithm. The peaks in the region $419-421$ nm 
previously unresolved are now distinguishable due to the SR algorithm. The physical reality of these separated peaks was verified by comparing the data obtained by the SR algorithm applied to our Ar data with that reported in NIST atomic line data reference. The difference in position 
between these values was a remarkable $0.01$ nm.  
The improvement in resolution, in terms of the cutoff frequencies, was extended from 
$\nu_c= 14.37 \times 10^8$ cm$^{-1}$ to  $\nu_\text{hi}= 10^{12}$ cm$^{-1}$. This is about 
3 orders of magnitude  or in terms of the super-resolution factor $SRF$ about $70$ times. In practical purposes this implies that the linewidth of a peak is reduced about $70$ times as compared to the original measured peaks.
This means, equivalently,  a percentage improvement of the signal (PEIS) of $98.5 \%$. 
 As a final remark, is important to emphasize that the reason why the peaks in the 
region $419-421$ nm can be well resolved by the SR algorithm is because the distance 
$\Delta$ between these peaks is about $0.3$ nm  and $2/\nu_c=1.4 \times 10^{-2}$ nm. 
Therefore, the minimal separation condition \eqref{minsep} holds and the hypothesis of the 
SR theorem~\cite{Candes2}  are satisfied.  Hence, up to the noise to signal ratio, 
the obtained estimate of the peak locations of the source signal is exact.\\
We showed in this work that dramatic results in resolution improvement for optogalvanic spectroscopy data in Argon using a Super-resolution algorithm. 
The results show an improvement in the resolution of the experimentally obtained data by a factor of $70$. The physical reality of the unraveled structure was checked 
by comparing the super-resolved spectrum with atomic transitions reported in the NIST data base. 
We point out that the identification of the appropriate filter $G$ that models accurately the experiment and the estimation of its parameters are two of the main issues to be address in order to correctly use of SR algorithms. 
We believe that SR algorithm could have a major impact on the analysis of experimental spectroscopic data. This is because it makes possible to obtaining high resolution spectral features without resorting to expensive and sophisticated experimental infrastructure. 

\acknowledgments
This project was supported by CONACYT CB$2011-167631$ and by PAPIIT DGAPA UNAM IN $106213$. Hoyos-Campo thanks CONACYT scholarship at PCF-UNAM. We would like to thanks the ICF-UNAM for providing the laser NT$342$B.

\end{document}